\documentclass[sigconf]{acmart}
\usepackage{algorithm}
\usepackage[noend]{algpseudocode}
\usepackage[utf8]{inputenc}
\usepackage{url}
\usepackage{graphicx}
\usepackage{enumitem}
\usepackage{xcolor}
\usepackage{amsmath}
\usepackage{caption}
\usepackage{subcaption}
\usepackage{booktabs}
\usepackage{afterpage}
\usepackage{floatpag}
\newcommand\sarah[1]{}
\newcommand\marko[1]{}
\newcommand\LRA{LRA}

\usepackage{algpseudocode}
\algdef{SE}[EVENT]{Event}{EndEvent}[1]{\textbf{upon event}\ #1\ \algorithmicdo}{\algorithmicend\ \textbf{event}}%
\algtext*{EndEvent}

\algdef{SE}[PARAM]{Param}{EndParam}[1]{\textbf{Parameters:}  #1}{\algorithmicend\ \textbf{parm2}}%
\algtext*{EndParam}

\algdef{SE}[IMPORT]{Import}{EndImport}[1]{\textbf{import} #1}{\algorithmicend }%
\algtext*{EndImport}

\algdef{SE}[INIT]{Init}{EndInit}[1]{\textbf{Init:} #1}{\algorithmicend }%
\algtext*{EndInit}

\newif\ifsubmission\submissionfalse
\newif\ifsa\safalse
\newif\ifproceeding\proceedingfalse

\title{Pikachu: Securing PoS Blockchains from Long-Range
Attacks by Checkpointing into Bitcoin PoW using Taproot}
  \author{Sarah Azouvi}
 \affiliation{%
 \institution{Protocol Labs}
  \country{}
 }
    \author{Marko Vukoli\'c}
 \affiliation{%
 \institution{Protocol Labs}
  \country{}
 }
\date{}

\copyrightyear{2022}
\acmYear{2022}
\setcopyright{rightsretained}

\acmConference[ConsensusDay '22]{Proceedings of the 2022 ACM Workshop on Developments in Consensus}{November 7, 2022}{Los Angeles, CA, USA}
\acmBooktitle{Proceedings of the 2022 ACM Workshop on Developments in Consensus (ConsensusDay '22), November 7, 2022, Los Angeles, CA, USA}
\acmDOI{10.1145/3560829.3563563}
\acmISBN{978-1-4503-9879-4/22/11}

\usepackage{etoolbox}
\makeatletter
\patchcmd{\maketitle}{\@copyrightpermission}{
   \begin{minipage}{0.3\columnwidth}
     \href{https://creativecommons.org/licenses/by/4.0/}{\includegraphics[width=0.90\textwidth]{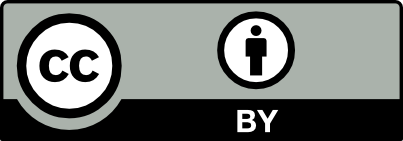}}
   \end{minipage}\hfill
   \begin{minipage}{0.7\columnwidth}
     \href{https://creativecommons.org/licenses/by/4.0/}{This work is licensed under a Creative Commons Attribution International 4.0 License.}
   \end{minipage}

   \vspace{5pt}
}{}{}

\makeatother

\begin{document}

\newcommand{\pk}{pk}
\newcommand{\sk}{sk}
\newcommand{\tx}{\textsf{tx}}

\makeatletter
\newcommand\xleftrightarrow[2][]{%
  \ext@arrow 9999{\longleftrightarrowfill@}{#1}{#2}}
\newcommand\longleftrightarrowfill@{%
  \arrowfill@\leftarrow\relbar\rightarrow}
\makeatother

\newcommand{\DKG}{\textsf{KeyGen}}
\newcommand{\abort}{\textsf{abort}}
\newcommand{\amount}{\textsf{amount}}
\newcommand{\RB}{\text{RB}}
\newcommand{\opreturn}{OP\_{RETURN}}
\newcommand{\ckpt}{\textsf{ckpt}}
\newcommand{\release}{\textsf{release}}
\begin{abstract}
   Blockchain systems based on a reusable resource, such as proof-of-stake (PoS), provide weaker security guarantees than those based on proof-of-work. Specifically, they are vulnerable to long-range attacks, where an adversary can corrupt prior participants in order to rewrite the full history of the chain. 
   To prevent this attack on a PoS chain, we propose a protocol that checkpoints the state of the PoS chain to a proof-of-work blockchain such as Bitcoin. Our checkpointing protocol hence does not rely on any central authority.
   Our work uses Schnorr signatures and leverages Bitcoin recent Taproot upgrade, allowing us to create a checkpointing transaction of constant size. We argue for the security of our protocol and 
   present an open-source implementation that was tested on the Bitcoin testnet.
\end{abstract}

\ccsdesc[500]{Security and privacy~Cryptography}
\keywords{Blockchain, proof-of-stake, long-range attack}

\maketitle

\section{Introduction}
\begin{figure}
    \centering
    \includegraphics[width = \linewidth]{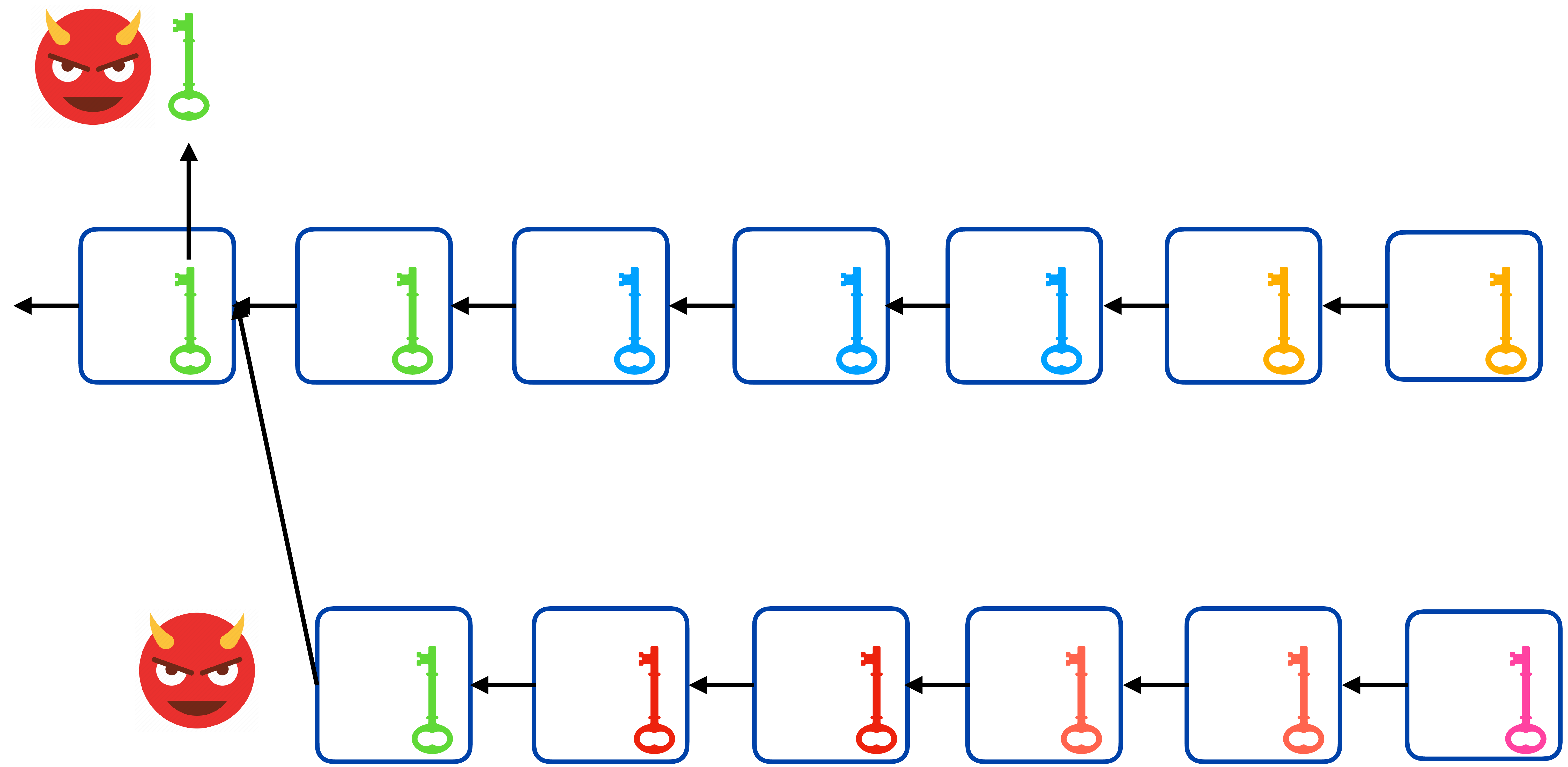}
    \caption{Illustration of the long-range attack. After the green validators (i.e., validators associated with the green key on the figure) left the system, the adversary acquired their keys. In a PoS blockchain, having access to validators' keys is enough to create new blocks and hence the adversary can create a chain as long as the honest chain (perhaps even simulating configuration change in its chain). Any user that trusted the green key and is presented with both chains cannot differentiate the honest from the adversarial chain.}
    \label{fig:LRA}
\end{figure}

Long-range attacks (\LRA) --- also called posterior corruption attacks~\cite{deirmentzoglou2019survey} ---  are one of the major security issues affecting permissionless proof-of-stake (PoS) blockchains. 
These attacks rely on the inability of a user who disconnects from the system at time $t_1$ and reconnects at a later time to tell that validators who were legitimate at time $t_1$ and left the system (by e.g., transferring their stake to other validators, or to themselves under a different identity) are not to be trusted anymore. 
In a PoS system,  where the creation of blocks is costless (i.e., does not cost physical resource such as energy), and timeless (i.e., is not rate-limited in time), 
these validators could create a fork that starts from the past, i.e., at time $t_1$, and runs until the present.
This is in sharp contrast to proof-of-work (PoW) systems, where creating blocks requires time (e.g., due to Bitcoin \cite{nakamoto2008peer} difficulty adjustment) and physical resources (e.g., energy for performing actual computation) and not just using cryptographic keys.
A client of a PoS blockchain would be unable to recognize the attack as they are presented with a ``valid'' chain fork. 
See Figure~\ref{fig:LRA} for a visual explanation of the attack.

Recently, Steinhoff et al.~\cite{steinhoff2021bms} proposed an approach to deal with \LRA\ by anchoring (checkpointing) the PoS membership into Ethereum’s proof-of-work blockchain (Eth 1.0), which is not vulnerable to this type of attack. The main idea of their work is to have a smart contract on the Ethereum blockchain that keeps track of the state of the membership of the underlying PoS system.
For a typical Byzantine Fault-Tolerant (BFT) protocol underlying a PoS blockchain, the smart contract on Ethereum would only be updated if, e.g., two thirds of the current staking power (or blockchain members in case of uniform voting rights) instruct the smart contract to do so.
In the approach of Steinhoff et al. each validator will send a transaction to the smart contract 
that indicates a vote for a new set of validators. As soon as two thirds of the votes for the same set have been received, the smart contract automatically updates its state to the new set.
From this moment on, the members of the new set are in charge of voting for the next set and so forth.
Every user that needs to verify that a set of validators are indeed legitimate and most recent ones, can do so by simply checking the smart contract. An adversary cannot change the state of the smart contract, even with the keys of
former validators, without creating a fork on the PoW blockchain, which is considerably more, if not prohibitively expensive. Any user can resort to the Ethereum smart contract to verify the correct state of the checkpointed PoS chain, effectively preventing the \LRA\ attack. 

However, as it happens, Ethereum is abandoning PoW and transitions to PoS \cite{EthereumMovingPoS} (Eth 2.0). Hence, the approach of Steinhoff et al. is no longer viable as PoS of Eth 2.0 cannot be used instead of PoW for anchoring as it is itself susceptible to the \LRA\ vulnerability. In this paper, we design a solution to \LRA, inspired by Steinhoff et al., using Bitcoin’s PoW, assuming that Bitcoin will never change its underlying consensus mechanism. The history of altcoin forks off Bitcoin and the Bitcoin development ethos give very realistic assurance that this assumption will hold.\footnote{The discussion on long-term viability of energy consumption of Bitcoin is out of scope of this paper and is available elsewhere \cite{Vukolic21}.}

However, the implementation and design of such a scheme on Bitcoin is more challenging, compared to the implementation of Steinhoff et al. on Eth 1.0, because Bitcoin’s scripting language expressivity is considerably more limited compared to smart contracts on Ethereum. Besides, the approach designed by Steinhoff et al. leverages multi-signatures for anchoring, which can quickly bloat the transaction size, making it at worst impossible to anchor PoS networks with large number of validators, or, at best, very costly to do so. 

\begin{figure}
    \centering
    \includegraphics[width = \linewidth]{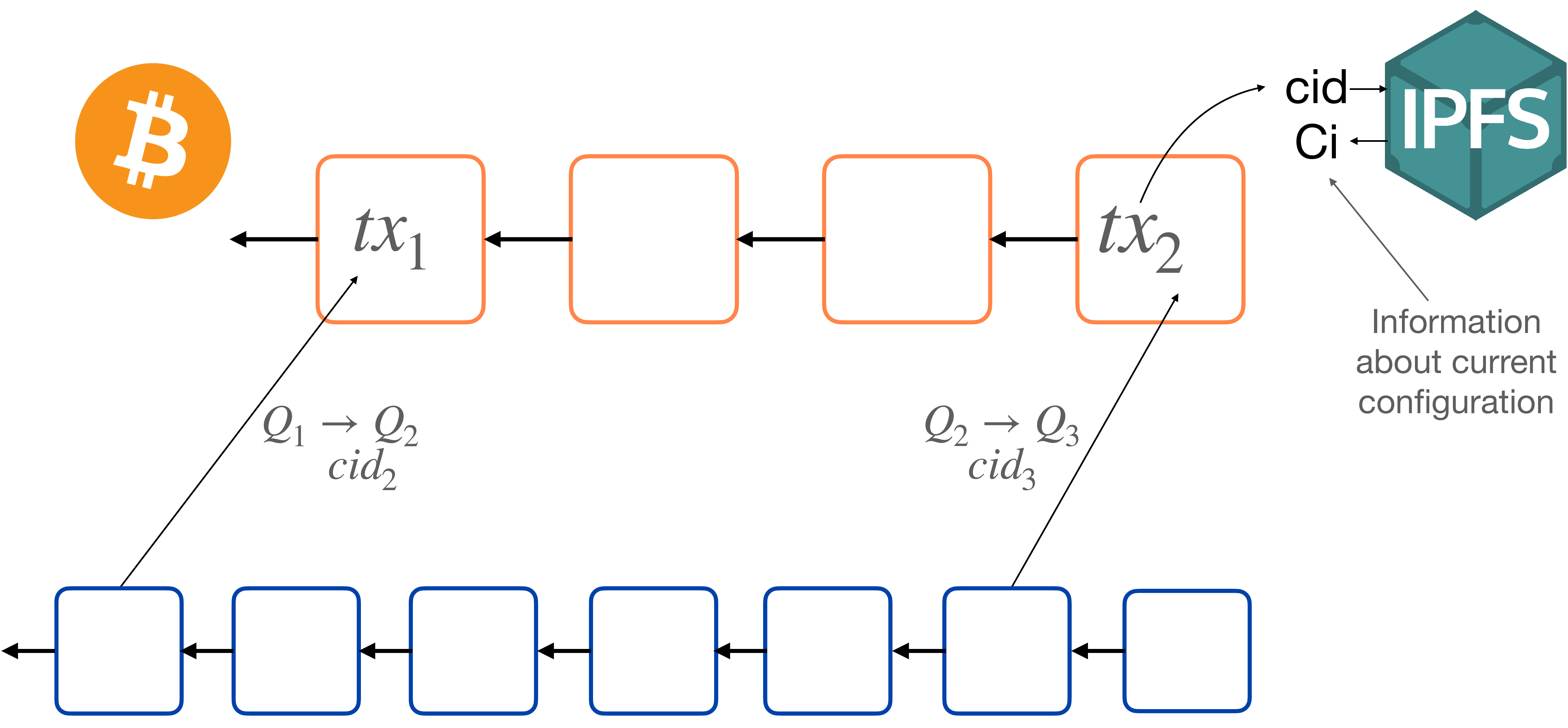}
    \caption{High-level visualization of the Pikachu protocol. Checkpoints from the PoS chain (in blue) are periodically pushed to the Bitcoin blockchain (in orange) by the PoS Validators. The checkpoints contain the Taproot address $Q$ (which itself contains the aggregated public key of the configuration and commitment to the PoS chain $\ckpt$) as well as a content identifier $cid$ that can be used with any content-addressable storage to retrieve information about the configuration (IPFS pictured).}
    \label{fig:pikachu}
\end{figure}

To address these limitations, our  approach is to use the capabilities enabled by the recent Taproot upgrade~\cite{bip341} to Bitcoin, which allows for more efficient Schnorr threshold signatures.
Briefly, our protocol, called Pikachu, works as follows.
As Bitcoin does not allow for stateful smart contracts, we use an aggregated public key to represent the configuration of validators $C_i$ in the PoS system. When the set changes significantly enough to configuration $C_{i+1}$, 
the aggregated key must be updated in the Bitcoin blockchain. This is done by having a transaction transferring the funds associated with the aggregated key of the previous validators $C_i$ to the new aggregated key controlled by validators in configuration $C_{i+1}$.
Instead of having each validator in $C_i$ send a transaction to the Bitcoin network,
this transaction is signed interactively, off-chain, and all the signatures are aggregated into one constant-size signature. Furthermore, we store the Merkle root of the state of the checkpointed PoS blockchain in the Bitcoin $\opreturn$ field of the transaction from $C_i$ to $C_{i+1}$. We store the data pertaining to this checkpoint off Bitcoin blockchain. While the data pertaining to the checkpoint could be stored anywhere (e.g., IPFS~\cite{ipfs}) and validated against the state root stored in the Bitcoin transaction --- our implementation uses a content-addressable key-value store implemented on top of the PoS system to store the actual checkpointed state. Figure~\ref{fig:pikachu} illustrates the high-level protocol.
We note that since our work is based on Schnorr threshold signatures and uses Bitcoin's Taproot, it could be of independent interest to any project looking to implement  threshold signing transactions on Bitcoin (for example, sidechains~\cite{bitcoinpeg}).

To summarize, our contribution is as follows. Starting from the observation that PoW gives much stronger security guarantees than PoS, we present a protocol to protect current PoS blockchains against \LRA\ by anchoring their state onto Bitcoin's blockchain. The advantage of using Bitcoin unlike, for example, a website, is that it is itself decentralized, hence our protocol does not add any single point-of-failure to a decentralized PoS system. We implemented our protocol on top of a delegated PoS blockchain and tested it on Bitcoin testnet storing checkpoints into a key-value store maintained by the PoS validators (although alternative storage method, such as IPFS could be used).

The rest of this paper is organized as follows.
We start by providing the necessary background in Section~\ref{sec:background} and our model and assumptions in Section~\ref{sec:assumptions}. We present our design in Section~\ref{sec:protocol} then a security argument in
Section~\ref{sec:security}.
Section~\ref{sec:implementation} presents the implementation of the protocol. We discuss related work in Section~\ref{sec:related}.

\section{Background}\label{sec:background}
We use elliptic curve notation for the discrete logarithm problem. Suppose $q$ is a large prime and $G,J$ are generators of a subgroup of order $q$ of an elliptic curve $\mathbb{E}$. We assume that $\mathbb{E}$ is chosen in such a way that the discrete logarithm problem in the subgroup generated by $G$ is hard, so it is infeasible to compute the integer $d$ such that $G = dJ$.

Let $H, H_1, H_2, H_{TapTweak}$ be cryptographic hash functions mapping
to $\mathbb{Z}_q^*$. 
We denote by $x\xleftarrow{\$}S$ that $x$ is selected uniformly at random from $S$.

\subsection{Schnorr signature}
The Schnorr signing scheme \cite{Schnorr91} works as follows.
Let $(s,Y)\in \mathbb{Z}_q^*\times \mathbb{E}$ be a user key pair (such that $Y=sG$) and $m$ a message to be signed. The signer performs the following steps.
\begin{enumerate}
    \item $k\xleftarrow{\$} \mathbb{Z}_q^*$
    \item $R\gets kG$
    \item $z \gets k+ H(m||R||Y)\cdot s \mod q$
\end{enumerate}
The signature is then $(z,R)$
and is verified by checking that $zG \stackrel{?}{=} R + H(m||R||Y)Y$.
\subsection{Secret sharing schemes}

A secret sharing scheme allows one participant (a dealer) to share a secret with $n$ other participants, such that any $t$ of them can recover the secret but any set of $t-1$ or less of them cannot.
Furthermore, a desirable property of a secret sharing scheme is to be publicly verifiable, i.e., anyone should be able to verify that the dealer computed the correct shares and did not cheat.
In this paper, we will use Feldman's  verifiable secret sharing scheme~\cite{feldman1987practical} (VSS), which we describe in
steps \textbf{1}-\textbf{3} of Figure~\ref{alg:dkg}.




\subsubsection{Generating a secret}
Unlike Feldman's VSS scheme, in which only one participant generates a secret and shares it with their peers, we consider a protocol where everyone contributes equally to generate a common secret, such that no set of participants of size strictly smaller than $t$ can recover the secret on their own.
We will use the scheme designed by Gennaro et al.~\cite{gennaro2007secure} that we define in Figure~\ref{alg:dkg}, and
we adopt the following notation:
$$(s_1,\cdots,s_n)\xleftrightarrow{(t,n)}
(r,Y,a_kG,S_0),\ k\in\{1,\cdots,t-1\}$$
to mean that $s_j$ is player $j$'s share of the secret $r$ for each $j\in S_0$.
The values $a_kG$ are the public commitments used to verify the correctness of the shares and $(r, Y )$ forms a key pair where $r$ is a private key and $Y$ is the corresponding public key.
The set $S_0$ denotes the set of players that have not been detected to be cheating during the execution of the protocol.
This protocol is secure for any $t>\frac{n}{2}$ (i.e., it can tolerate an adversary that corrupts up to half of the participants).

\subsection{Threshold signing}
A $t$-of-$n$ threshold signing scheme allows any combination
of $t$ participants to sign a message while preventing any coalition of $t-1$ participants or less to create a valid signature, i.e., at least $t$ participants must agree to sign the message for the signature to be valid.
We use the threshold signing protocol FROST~\cite{komlo2020frost}, that we define in Figure~\ref{alg:signing}.
This interactive protocol will either output
a Schnorr signature $(z, R)$ on a message $m$ or a $\abort$ message, together with a set of misbehaving participants such that the protocol can be rerun without these misbehaving participants in the next step.
The protocol relies on a signature aggregator (SA),
however, as the main role of the SA is to choose the subset of participants designated for signing, it can easily be removed. Instead, we can have each participant compute the set in a deterministic way.
Alternatively, in the case of PoS chain, we could choose this set pseudo-randomly using some randomness coming from the chain (random
numbers are often created as part of a PoS protocol as they are needed for, e.g., leader
election). 

\paragraph{Choice of the Schnorr signing protocol.} We chose to use the FROST signing protocol because it is more efficient than alternative protocols, such as the Stinson and Strobl~\cite{stinson2001provably} protocol, even though it is not robust, i.e., the protocol cannot complete if one participant aborts or misbehaves. However, misbehaving participants are detectable in FROST (each public share is verifiable against a public key), so the protocol can simply be restarted from scratch without those malicious participants.
We did not use other Schnorr signing protocols~\cite{nick2020musig2,drijvers2019security} as they are not compatible with threshold signing.

Note that we did not implement the key generation algorithm presented by Komlo and Goldberg~\cite{komlo2020frost}, used originally in FROST, as it does not allow to detect misbehaving participants, therefore losing the ability to re-start the protocol without the misbehaving participants.
Instead, we will use the scheme by Gennaro et al.~\cite{gennaro1999secure} and borrow only the signing scheme presented in the FROST paper~\cite{komlo2020frost}, as per the authors' suggestion.
The distributed key generation (DKG) algorithm by Gennaro et al. is also used by Stinson and Strobl~\cite{stinson2001provably} and has the advantage of being robust (it will complete despite misbehaving participants, who are detected through a complaint process). We follow the suggestion in Gennaro et al.~\cite{gennaro2007secure}  and use the simpler variant of the DKG, \textsf{JF-DKG}, as this is sufficient for our application of threshold signing.

The main reason for preferring an efficient but non-robust signing algorithm is that our protocol will eventually be incentivized (financial rewards will be given out to participants who perform the signature). Therefore, it is reasonable to expect participants to cooperate, especially when malicious behavior is detectable and can only delay --- not prevent --- the signing.
Because both the DKG and the signing part of our protocol are modular, other threshold signing protocols can be used interchangeably for different threat models (e.g., including the robust signing protocol in~\cite{stinson2001provably}).

\subsection{Taproot}
Taproot is a recent Bitcoin network upgrade  that allows transactions to be signed using Schnorr signatures and that introduces a new data-structure, Merkelized Abstract Syntax Trees (MAST), for more advanced scripting in a privacy-preserving way.
The main advantage of Schnorr signatures over the ECDSA multi-signature is that they enable signature aggregation, saving space in Bitcoin blocks while also providing more privacy as it is not possible to distinguish between a ``regular'' transaction, i.e., sending bitcoins from one person to another, and a more complex one, e.g., using a threshold signature. This could help hide identities in the blockchain and thwart clustering deanonymization~\cite{meiklejohn2013fistful} although we are not interested in this property for this work.

\begin{figure}[h]
    \centering
    \includegraphics[width = \linewidth]{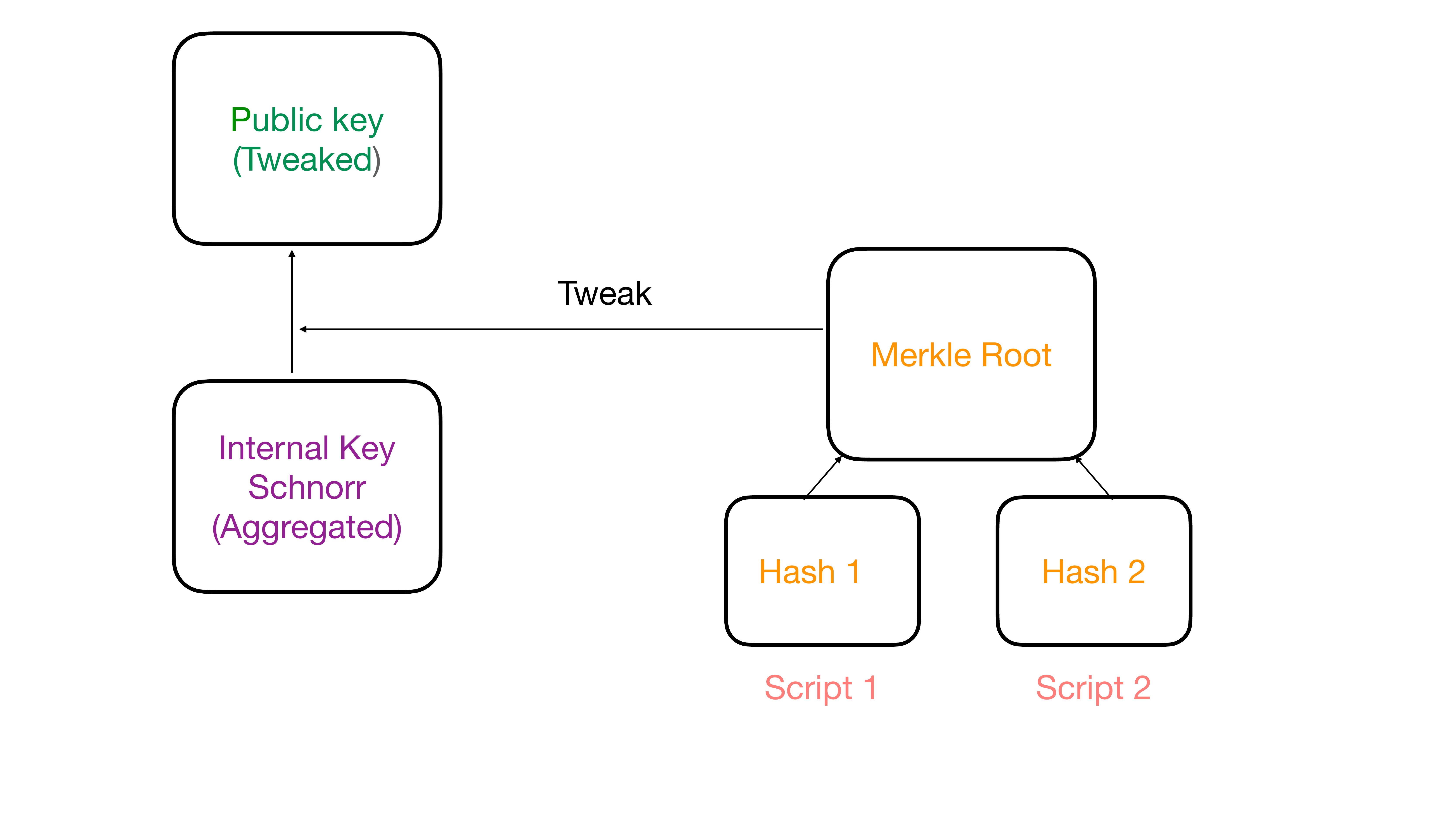}
    \caption{Taproot Output Composition}
    \label{fig:taprootaddress}
\end{figure}

A Taproot address has two components:
a single public key (the internal key) and a script tree, identified by its Merkle root.
Either component can be used independently to spend the UTXO.
In the case of threshold or multi-signatures, the internal key can be the aggregated public key of all the signers.
The script tree can contain an arbitrary number of different scripts, each of which specify a condition that must be satisfied in order for the coins to be spendable.
For example, one condition can be to give the pre-image of a hash.
As the name suggests, in the script tree, the scripts are organized in a tree (see Figure~\ref{fig:taprootaddress}). The transaction can be spent either by using the internal secret key (key path) or by satisfying one of the conditions in the tree (script path).
In this paper, we are interested in spending a Taproot output using the key path.
It should be noted that it is possible to use the script tree to define a threshold signature scheme~\cite{murch_2020}, though less efficient as the size of the tree would grow exponentially with the number of participants~\cite{bitcoinpeg}.

We now detail how to spend a Taproot output using the key path.

\subsubsection{Key path spending}
To prevent a potential vulnerability in which one user of a threshold or multi-signature could steal all the funds~\cite{bitcoin_2021}, the output key should commit to a (potentially unspendable) script path even if the spending condition does not require a script path (i.e., if only the key path is going to be used).
There are multiple ways to achieve this with Taproot.
The most natural way is to simply include the internal public key in the ``tweak.''
The tweaked public key (i.e., outer key) is then computed as follows:
$$Q = P + \text{int}(H_{TapTweak}(bytes(P)))G$$ where
$P$ is the internal public key and
$H_{TapTweak}$ is a hash function.
The associated tweaked private key is then:
$$q = p + \text{int}(H_{TapTweak}(bytes(P)))$$ where $p$ is the private key associated with $P$. In order to spend the output using the key path, one must then sign the transaction with the tweaked private key.

\paragraph{Adding a commitment}\label{sec:background-taprootcommitment}
Alternatively, the script path could be used to add a commitment. For example in our case this commitment could be the hash of the underlying PoS chain at regular intervals.
Let $c$ denote this commitment.
In this case, the tweaked public key becomes: 
$Q = P + H_{TapTweak}(P||c)G$ and the tweaked private key 
$q = p + H_{TapTweak}(P||c)$.
The script path is still unspendable, and the output is spent by signing using the tweaked private key.

\subsubsection{Transaction notation}
For any Bitcoin transaction, we use the following notation:
$$\textsf{input}_1,\dots,\textsf{input}_i \rightarrow ((\amount_1,\textsf{output}_1),\dots,(\amount_j,\textsf{output}_j))$$ to say that all the coins associated with $\textsf{input}_1\dots,\textsf{input}_i$ are transferred to $\textsf{output}_1,\dots,\textsf{output}_j$ with, respectively,
$\amount_1$,$\dots$, 
$\amount_j$.
As a reminder, since Bitcoin is UTXO based, all the coins from an input must be transferred during the transaction, although to potentially multiple addresses.
Additionally, it must be the case that $\amount_1+\cdots+\amount_j\le  \textsf{input}_1.\amount + \cdots +\textsf{input}_i.\amount$
where $\textsf{input}_k.\amount$ represents the total amount associated with $\textsf{input}_k$.
The remaining amount (in the case of a strict inequality) is used as a transaction fee for the miner mining the block.


\section{Model and Assumptions}\label{sec:assumptions}
We assume an underlying blockchain based on a reusable resource such as PoS \ifsubmission\else{or proof-of-space }\fi.
Each state of the PoS blockchain is associated with a set of participants, called the configuration and denoted by $C$, and their corresponding \emph{power}  \ifsubmission{(e.g., number of coins staked in the case of the PoS).}\else{(e.g., number of coins staked in the case of proof-of-stake and storage space in the case of proof-of-storage).}\fi
We call the set of weighted participants in a configuration the \emph{power table}.
The power table is determined by a set of signing keys and their associated weight:
$C=\{(PoS.pk_i,w_i)\}_{i=1}^{|C|}$.
Each signing private key $PoS.sk_i$ is private to
$i$-th participant. 
For simplicity, we consider a flat model, i.e., one participant accounts for one unit of power in the PoS blockchain, hence we omit the weight from our model moving forward.
The flat model could be generalized by considering that one participant with $x$ units of power possesses $x$ public keys, one for each of their units of power. We will discuss how this assumption impacts the scalability of our protocol in Section~\ref{sec:open-problems}.
Furthermore, we assume that there is some similarity between successive configurations of the system, i.e., the set of participants does not change completely from one configuration to another.
Formally, 
we define the difference between two configurations
$C_j$ and $C_i$ as their symmetric difference ($C_i\triangle C_j$), which
corresponds to the number of reconfiguration requests that
need to be applied to $C_i$ in order to obtain $C_j$. 
We assume that for two consecutive configurations $C_i$ and $C_{i+1}$,
their symmetric difference is bounded by some parameter $b$.

Following~\cite{azouvi2020winkle}, we define a \emph{perpetually honest} participant as a participant that follows the protocol and maintains the secrecy of their
signing keys in perpetuity (an adversary may never have access to
them). This is opposed to an \emph{eventually compromised} participant who after some time, leaks all its previous
signature keys to the adversary.

We assume that the PoS is secure, i.e., satisfies the usual security properties of consistency, chain growth, and chain quality~\cite{garay2015bitcoin}, as long as a sufficient fraction of the participants are perpetually honest.
Let $f$ be the maximum fraction of power that an adversary can control
 while the protocol maintains its security when the rest of the power table is perpetually honest (e.g., $f =1/3$).
For simplicity, we assume that this blockchain provides instant finality, i.e., that there  are no forks.
This can be achieved using some variant of a BFT-protocol~\cite{gilad2017algorand,buchman2016tendermint} or relaxed by using a ``lookback'' parameter. For example, if a block is final after $k$ confirmations, then we will use the state of the chain $k$ blocks in the past instead of the latest state to ensure consistent views across participants.

For the rest of this paper we will consider the security of the PoS chain
under eventually compromised honest participant as follows.
We consider an adversary $\mathcal{A}$ that, for each state $i$ of the PoS system, controls \textit{all the keys} from previous configurations $(C_j)_{j<i-L}$ where $L\gg 1$ is a parameter (assumption 1) as well as a fraction of at most $f$ participants in configurations $(C_j)_{i-L\le j\le i}$ (assumption 2). We quickly note that $f<\frac{1}{2}$ since there does not exist any protocol that is secure with $f>\frac{1}{2}$.

Under this assumption, the adversary is able to mount a \LRA\ as follows.
The adversary starts a fork of the PoS chain at height $j<i-L$, using the keys from configuration $C_{j}$ and that runs until the current height $i$. Since the adversary does not hold the keys from configuration $i-L$ and above, this means that from this height, the configurations on the adversarial fork and on the honest chain must differ.
Note that under this attack, any online validator is able to differentiate the correct chain from a chain created as part of a \LRA\ (since they are not part of the configurations in the adversarial fork).
In the rest of the paper we use \emph{correct chain} to
mean the chain in the view of the online validators.
Our protocol will ensure that any user is also able to distinguish each chain even if they have been offline, by looking at the Bitcoin blockchain.
We discuss the security properties that the protocol should achieve in Section~\ref{sec:security}. 

We add another, optional, assumption: the existence of a random beacon $(\RB_i)_{i\in\mathbb{N}}$ that emits a new randomness for each state of the database (i.e., at each height of the underlying PoS blockchain).
This is a standard assumption in PoS blockchains as a random beacon is necessary for the leader election part of the protocol. 
This randomness will be used by participants to pseudo-randomly select the set of signers. Another option would be to select this set in any deterministic manner.

Lastly, participants will use the PoS chain to broadcast the messages relative to our Pikachu protocol (although another broadcast channel could be implemented alternatively). We assume that each message is included in the chain (or broadcast) after a small number of blocks.
\section{Protocol}\label{sec:protocol}

\subsection{Overview}
The intuition behind the protocol is as follows:
each configuration $C_i$ is associated with a Taproot public key $Q_i$ that consists of an internal key, in this case an aggregate public key $\pk_i$, that participants computed with an interactive DKG protocol (step~\ref{enum:init} of the main algorithm protocol in Figure~\ref{alg:main}) and a tweaked part as defined in Section~\ref{sec:background-taprootcommitment}.
 We chose to tweak the internal key using a commitment to the PoS chain (i.e., the hash of the state of the PoS blockchain).
Each player $j$ in the configuration then knows a share of the secret key associated with $\pk_i$, $s_{i,j}$, such that $t_i$ of the shares are enough to compute a valid signature on any message, but fewer than $t_i$ participants cannot compute a signature.
Configuration $C_i$ is responsible for anchoring the state of the PoS chain at this point in time in the Bitcoin blockchain, which also includes updating the new configuration.
In order to do so, the new configuration $C_{i+1}$ must first compute their aggregated public key $\pk_{i+1}$ using the DKG algorithm.
This key is then tweaked using a commitment $\ckpt$ to the PoS chain (i.e., the hash of the PoS chain at that time). The tweaked key becomes $Q_{i+1} = \pk_{i+1} + H_{TapTweak}(\pk_{i+1}||\ckpt)G$.
Note that only the tweaked key will appear on the blockchain so the hash $\ckpt$ will not be visible by anyone looking at the blockchain without external knowledge. However, anyone who has access to $\pk_{i+1}$ and $\ckpt$ can easily reconstruct $Q_{i+1}$ to verify that their view of the PoS chain is correct.

To update the configuration from $C_i$ to $C_{i+1}$, a transaction from $Q_i$ to $Q_{i+1}$ must be included in the Bitcoin blockchain (steps~\ref{enum:signing} and \ref{enum:taproot} in Figure~\ref{alg:main}).
Leveraging the recent Bitcoin Taproot upgrade (that allows for Schnorr signatures), the transaction needs to be signed by $t_i$ participants from configuration $C_i$
where $t_i$ is chosen to be strictly more than $f|C_i|$ as
this ensures that at least one honest participant signs, preventing an adversary from signing an illegitimate transaction. As discussed previously, we will use the FROST algorithm for signing.
Note that, the DKG requires that $t_i> 0.5|C_i|$ to ensure security so
our final constraint on $t_i$ is $t_i> \max(0.5|C_i|,f|C_i|)$.
Since we assume that online validators can distinguish a \LRA\ chain, it is enough to have the transaction signed by $t_i$ participants as no honest validators can be fooled into signing an illegitimate transaction.
If forks were allowed even in the case of perpetually honest validators (i.e., outside of \LRA\ forks), this would be more problematic, as two conflicting transactions could then be signed, and we would require at least two thirds of the participants to sign the transaction, for $f=1/3$ (as previously mentioned, this can also be fixed by considering a block in the past, i.e., one that has been finalized).

In addition to the transfer of coins from $Q_i$ to $Q_{i+1}$, the transaction spent by configuration $C_i$ will have a second output that does not receive any bitcoins and that is unspendable, but that contains an identifier $cid$ used to retrieve the full details of the configuration.
This is done using the $\opreturn$ opcode of Bitcoin~\cite{opreturn} that allows storing of extra information in the chain, which we use to store $cid$.
This identifier will be useful in the case where a user does not have access to the right PoS chain (i.e., does not have the correct value for $\pk_{i+1}$ and $\ckpt$ due to a \LRA).
In this case, the content identifier $cid$ can be used, together with a content-addressable decentralized storage, for example IPFS~\cite{ipfs} or Filecoin~\cite{filecoin} (or a content-addressable storage implemented on the PoS network validators) to retrieve the identities of the nodes in the correct configuration.
The transaction updating the configuration will look as follows:
$$\tx_i:Q_i\rightarrow((\textsf{amount},Q_{i+1}),(0,\opreturn=cid_{i+1}))$$ meaning that $\textsf{amount}$ is transferred to $Q_{i+1}$ and 0 is transferred to $\opreturn=cid_{i+1}$ (unspendable output).
This information is then publicly available.
We discuss in Section~\ref{sec:verify} how any user can then use it to get the latest PoS configuration.

We add the following assumption (assumption 3): we assume that $\tx_i$ is finalized in the Bitcoin blockchain before the configuration $C_{i+L}$ is formed, where $L\gg 1$ is the parameter defined in assumption 1 (Section~\ref{sec:assumptions}).

The high-level description of the protocol is presented in Figure~\ref{alg:main} and the pseudocode in Algorithm~\ref{pseudocode:main}. The pseudocode for our DKG and signing subroutines are presented in Algorithms~\ref{pseudocode:dkg} and ~\ref{pseudocode:signing}.
In all our pseudocode, the notation
$\langle msg \rangle_i$ means that message $msg$ was sent by participant $i$ and 
we use PM($\langle msg\rangle,i$) to denote that a private message $msg$ was sent to participant $i$.

We make the following remarks about our protocol. First, in steps~\ref{enum:push-ipfs} and~\ref{enum:send-tx-btc} we ask that every participant $P_j$ in configuration $C_{i}$ publishes the configuration state to the decentralized storage provider and sends the signed transaction $\tx_{i}$ to the Bitcoin network. We do so out of caution. In practice only one validator needs to do so, but this validator could be controlled by the adversary and abort instead.

Second, in step~\ref{enum:taproot} of the protocol, we remark that the final signature on the transaction, $z'$, is ``tweaked'' using $$H(\tx_i||R||Q_i)H_{TapTweak}(\pk_i||\ckpt).$$ This is because because the signature computed as part of the FROST signing algorithm  will verify against the key $\pk_i$, computed during the DKG but not $Q_i=\pk_i+H_{TapTweak}(\pk_i||\ckpt)$. For the signature to be valid on the taproot output, the signature must verify against the tweaked key $Q_i$. Because Schnorr is additive, it is enough to add the term $H(\tx_i||R||Q_i)H_{TapTweak}(\pk_i||\ckpt)$ to the signature.
Indeed one can verify that if 
$zG = R + H(\tx_i||R||Q_i)\pk_i$ then 
\begin{align*}
  z'G &= zG + H(\tx_i||R||Q_i)H_{TapTweak}(\pk_i||\ckpt)G  \\
  &= R + H(\tx_i||R||Q_i)\pk_i + H(\tx_i||R||Q_i)H_{TapTweak}(\pk_i||\ckpt)G\\
  &= R +  H(\tx_i||R||Q_i)(\pk_i +H_{TapTweak}(\pk_i||\ckpt)G)\\
  &= R +  H(\tx_i||R||Q_i)Q_i
\end{align*}

\begin{figure*}
    \centering
\noindent\fbox{%
    \parbox{\textwidth}{%
    {\small
We assume that the initial aggregated public key of participants (at genesis) $pk_0$ as well as their tweaked key $Q_0$  are trusted and known by everyone and that
it was funded as specified in Section~\ref{sec:init} such that there are enough bitcoins to pay for the transaction fees of several transactions.
For each round $i>0$:
\begin{enumerate}[label=\textbf{\arabic*}]
\item \label{enum:init} The protocol starts 
after a threshold of new registrations and unregistrations has been monitored (e.g., since the last configuration, $i$, there has been $u$ new registrations or unregistrations). We call this event $U_{i+1}$.
We note $X_{i+1}$ the height, in the PoS blockchain, corresponding to this event. 
As soon as the parties notice event $U_{i+1}$, they start the distributed key generation algorithm defined in Figure~\ref{alg:dkg}.
This algorithm is performed by members of the \textbf{new configuration}, $C_{i+1}$ in order to compute the new aggregated key $\pk_{i+1}$. We denote $S_{i+1,0}$ the set of members in the new reconfiguration (i.e., reconfiguration $i$).
(Every member knows who is part of the new configuration by property of the underlying PoS, using the power table). At the end of the algorithm, the aggregated public key $\pk_{i+1}$ is known by everyone and a message can be signed by $t_{i+1}$ out of $n_{i+1}$ of the participants using their secret share $s_{i+1,j}$:
$(s_{i+1,1},\cdots,s_{i+1,n})\xleftrightarrow{(t_{i+1},n_{i+1})}
(\sk_{i+1},\pk_{i+1},a_{i+1k}G,S_{i+1,1}),\ k\in\{1,\cdots,t_{i+1}-1\}$.
Here $S_{i+1,1} = S_{i+1,0} \setminus \{\text{misbehaving participants from the protocol}\}.
$

We assume that the DKG is finished by block $X_{i+1}+Y$ where $Y$ is chosen conservatively.
The tweaked public key of the taproot address is then defined to be
$Q_{i+1} = \pk_{i+1} + H_{TapTweak}(\pk_{i+1}||\ckpt)G$, where $\ckpt$ is the hash of the PoS block at height $X_{i+1}$.
\ifsubmission\else{\item Optional: Remove the misbehaving party from the power table.}\fi

\item \label{enum:signing} Signing protocol. Every participant $P_j$ of configuration $C_{i}$ does the following:
\begin{enumerate}

\item $P_j$ checks that the previous reconfiguration transaction $\tx_{i-1}$ (according to the PoS blockchain) is included in the bitcoin blockchain. If not, they submit it before forming the new  transaction.
\item\label{enum:push-ipfs} $P_j$ first publishes the list of members in the new configuration $C_{i+1}$ to the decentralized storage and retrieves the corresponding content identifiers $cid_{i+1}$. 
\item $P_j$ computes transaction $\tx_i$  as follows.
All of the coins associated with $Q_{i}$ are transferred to $Q_{i+1}$ and another output that receives no coins but contains an $\opreturn$ that contains $cid_{i+1}$ is added:
$\tx_i: Q_{i} \rightarrow ((\textsf{amt}, Q_{i+1}),(0,\opreturn = cid_{i+1}))$ where $\textsf{amount}$ is the amount associated with $Q_i$ minus transaction fees.

\item The members of the \textbf{current configuration} $C_{i}$
(i.e. associated with $\pk_{i}$)
perform the interactive signing algorithm.
\begin{enumerate}
\item Set $m\gets 0$.
\item \label{enum:sign} $(o, S_{i,m+1})\gets \textsf{SchnorrThresholdSign}(S_{i,m},\tx_i,\pk_i,Q_i)$ defined in Figure~\ref{alg:signing}
where $S_{i,m+1}$ is the set of non-misbehaving parties during the execution of the protocol.
\item If $o=(z,R)$, i.e., a signature has been successfully produced, continue to step~\ref{enum:taproot}.
\item Else (i.e., $o = \textsf{abort}$) set $m=m+1$ and go to step~\ref{enum:sign}.
\end{enumerate}
\end{enumerate}
\item \label{enum:taproot}The taproot signature is then computed as
$(z',R) \gets (z + H(\tx_i||R||Q_i)H(\pk_i||\ckpt),R)$, where $c$ is the hash of the PoS blockchain at height $X_i$.
\item\label{enum:send-tx-btc} $P_j$ sends  $\tx_i$ to the Bitcoin blockchain to update the configuration.
\item 
Participants set $i\gets i+1$ and go back to step~\ref{enum:init}.
\end{enumerate} }}}

    \caption{Main Algorithm}
    \label{alg:main}
\end{figure*}

\begin{figure*}
    \centering
\noindent\fbox{%
    \parbox{\textwidth}{%
    {\small
    Each participant $P_i$ performs the following steps, where $t$ is a parameter and $n$ is the total number of participants:
    \begin{enumerate}
        \item\label{enum:dkg-createsecret} Choose $r_i\xleftarrow{\$} \mathbb{Z}_q^*$.
        Let the sharing polynomial be $f_i(u)=\sum_{k=0}^{t-1}a_{ik}u^k$ 
        where $a_{i0}=r_i$.
        Compute $s_i^j =f_i(j) \mod q $ for each $j\in \{1,\dots n\}$ and send $s_i^j$ privately to $P_j$.
        \item Expose $Y_i = r_iG$ as follows. Broadcast $A_{ik} = a_{ik}G$ for $k\in\{ 0,\cdots,t-1\}$.
           \item\label{enum:dkg-verifyshares} Verify the values broadcast by other players: $f_j(i)G \stackrel{?}{=} \sum_{k=0}^{t-1}i^k A_{jk}$.
            If the check fails for an index $j$, complain against $P_j$.
\item\label{enum:complaint-answer} Answer each complaint from party $P_j$ against $P_i$ (if any) by broadcasting
$s_i^j$.
\item If any of the revealed shares fails this equation, remove that participant from the set of players $H_0$. 
\item Extract $Y = \sum_{j\in S_0} r_jG$, of which each player's share of the secret is $s_i = \sum_{j\in S_0} s_j^i$.
The secret $r= \sum_{j\in S_0} r_j\mod q$
is never computed.

    \end{enumerate}
The corresponding aggregated private and public keys are
$(r,Y)$, denoted by $$(s_1,\cdots,s_n)\xleftrightarrow{(t,n)}
(r,Y,a_kG,S_0),\ k\in\{1,\cdots,t-1\}$$
}}}
    \caption{Distributed Key Generation Algorithm ($\textsf{JF-DKG}$ by Gennaro et al.~\cite{gennaro2007secure})}
    \label{alg:dkg}
\end{figure*}

\subsection{Initialization and funding}\label{sec:init}
The initial key $Q_0$ is created by having the first configuration run the DKG, and tweak it with a hash of the genesis block of the PoS chain.
In order to fund the initial transaction, we want each participant in $C_0$ to send a small amount of bitcoins to $Q_0$. However it is not possible to enforce this. A participant that does not contribute to the fee would still hold a share of the secret key associated with $Q_0$.
Indeed $Q_0$ must be determined before the participants send their transactions, otherwise they do not know where to send their funds. But once $Q_0$ is computed everyone who participated in the DKG knows a share of the secret regardless of whether they send some funds to it.
We thus need to make sure that participants are incentivized to contribute to the fees.
One way to do so is to have each participant
who sent some funds to $Q_0$ in the Bitcoin blockchain be rewarded, in exchange, with some PoS coins. Verifying the validity of Bitcoin transactions is, however, not trivial.  Verifying the signature only is not enough as the transaction could be double spending. Hence, additional data is required by a verifier. More specifically, a verifier would need to verify that the transaction is included in the Bitcoin blockchain at least $k$ blocks deep - where $k$ is a parameter corresponding to Bitcoin's settlement time.
With this in mind, we propose the following protocol.

We consider the following parameters: a deadline $h_0$ (represented as a height in the Bitcoin blockchain); the settlement time $k$ after which a block is considered "finalized" in the Bitcoin blockchain (e.g. 6 blocks); $\release$ expressed as a height in the Bitcoin blockchain chain, chosen conservatively high.

\begin{enumerate}
    \item Each participant $P_i$ in $C_0$ submit a commitment to their Bitcoin public key $btc.pk_i$ (e.g. a hash $H_1(btc.pk_i)$) to the PoS chain. This is to prevent participants from later on "stealing" each other rewards by pretending to have sent some bitcoins that someone else sent. 
\item Configuration $C_0$ interactively performs the DKG to create the key $pk_0$. Each participant in $C_0$ holds a share of the secret key associated. The key is then tweaked with a commitment to the PoS genesis block to give $Q_0$.
\item  Each participant $P_i$ in $C_0$ send a small amout $\textsf{fee}$ from $btc.pk_i$ to $Q_0$. This transaction should be sent several heights before height $h_0$. They add a timelock~\cite{timelock} such that if the output is not spent after $\release$ blocks, $P_i$ gains control of their bitcoins back. We denote this transaction $init.tx_i$.
\item  Once the Bitcoin chain has reached height at least $h_0+k$ the participants can start the interactive signing. They create the transaction by spending all the UTXOs that were received by $Q_0$ before block $h_0$ (this ensures that everyone will sign the same transaction). We note $\tx_0$ this transaction. Every transaction $init.tx_i$ not included in the initial transaction $\tx_0$ (e.g. because it was included too late in the bitcoin blockchain) can be sent back to its original sender due to the timelock.
\item  If $init.tx_i$ was included in the inputs of $\tx_0$, $P_i$ can submit evidence of this in the Filecoin chain using $\tx_0$ (i.e. everyone can verify that $init.tx_i$ is in the list of input of $\tx_0$ and that the signature is correct). Since no adversary can forge a signature from $Q_0$, due to the security of the threshold signing scheme, no proof can be forged for $init.tx_i$ inclusion in $\tx_0$.
\item  If $init.tx_i$ was not included in the inputs of $\tx_0$, then $P_i$ does not get any reward.
\item  Every PoS miner verifies that $init.tx_i$ was indeed included in $\tx_0$ (as described above), then verifies that $btc.pk_i$ indeed belongs to $P_i$. If both checks pass, $P_i$ is awarded an amount of PoS coins proportional to the amount sent by $init.tx_i$. This amount should be high enough to not only compensate the fee paid  by $P_i$ but also incentivized them to sent the fee (i.e., the reward must be higher than the fee, although it is not trivial to compare the value of two cryptocurrencies, these values can be chosen conservatively). The reward can be taken from the coins minted, as is usually the case in crypto-currencies reward scheme. 
\end{enumerate}

For every checkpointing transaction on the Bitcoin blockchain, we use a constant fee $\textsf{btc.fee}$ chosen high enough to tolerate potential congestion period in the Bitcoin blockchain. As a reminder, thanks to the Taproot update, the size of the transaction in our protocol is constant in the number of participant hence choosing a constant transaction fee is enough for our purpose, although we may end up over-paying during non-congested periods.
We also remark that our protocol is assumed to be run at a relatively low pace (e.g., once a day) hence we can tolerate longer delays in having the checkpointing transaction included in the Bitcoin chain in periods of short-term congestion.
For reference , as of May 2022, the cost of a checkpointing transaction on Bitcoin mainnet would be around 
\$0.07 (around 200 sats).

When the funds from the initial transaction run out, a protocol
as the one described above can be used to refill them.

\subsection{Verification}\label{sec:verify}
Once the protocol described in Figure~\ref{alg:main} has been run by the participants, users of the PoS system who went offline for an extended period of time can use the Bitcoin blockchain to determine the correct configuration and state of the chain.
Informally, the verification protocol works as follows: users, who are aware of the initial aggregated public key $Q_0$, which serves as an identifier of the PoS blockchain on the Bitcoin blockchain, can follow the chain of transactions from $Q_0$ to the newest public key $Q_i$.
The latest transaction in the chain (i.e., from $Q_{i-1}$ to $Q_i$) contains an additional output that corresponds to the content identifier of the configuration $C_i$.
The user can then use this identifier to retrieve the configuration using IPFS (or another content-addressable decentralized storage, e.g., one implemented on top of the PoS chain).
The high-level protocol is described below and the pseudocode is given in Algorithm~\ref{pseudocode:verif}.

\begin{enumerate}
     \item Synchronize with the Bitcoin blockchain (e.g., by running a Bitcoin full node.\footnote{Bitcoin full nodes can be run on relatively cheap hardware, e.g., Raspberry Pi and 1TB disk, in a setup that costs less than \$200 USD.})
     \item Look for $Q_0$ and follow the chain of transactions to get $\tx_i$ and $cid_i$, i.e.,
     \begin{enumerate}
        \item Inspect the transactions going out from $Q_0$
        \item\label{enum:verif-init} If there are multiple transactions going out from $Q_0$, look for the initial funding transaction by inspecting the UTXOs spent and verifying that all of them are included in blocks with height lower than $h_0$. 
        \item Once the initial transaction $\tx_0$ has been found, look for the transaction that spent $\tx_0$ (i.e. where $\tx_0$ is an input).
        \item\label{enum:verif-induction} For $i\ge 0$ get $\tx_{i+1}$ by looking for the transaction that spent $\tx_i$.
        \item Stop when $\tx_i$ is unspent and get $cid_i$ from the $\opreturn$ field.
     \end{enumerate}
     \item Use $cid_i$ to get the list of current nodes from the external storage chosen.
     \item Request the PoS blockchain state from these nodes.
     \item \label{enum:verif} Verify that the aggregated public key on the PoS blockchain $\pk$ and the hash of the block $\ckpt$ are in accordance with the Bitcoin Taproot address $Q$ that is the output of $\tx_i$.
    
 
     
     
    \item If the checkpoint and aggregated key do not match
    the Bitcoin checkpoint, roll back the PoS chain until the previous checkpoint and go back to step~\ref{enum:verif}.
 \end{enumerate}

\begin{figure*}[h!]
    \centering
\noindent\fbox{%
    \parbox{\textwidth}{%
    {\small 
    $\textsf{SchnorrThresholdSign}(H,m,Y,Q)$\\
    Input: $H$ is the set of players,  $m$ the message.\\
    Y is the aggregated public key. Each participant $P_i$ holds a share of the associated secret key $s_i$. $Y_i$ is the public verification share of each participant and is computed as $Y_i = \sum_{j\in S_0}\sum_{k=0}^{t-1}i^kA_{jk}$
    We note $Q$ the tweaked key as defined in step~\ref{enum:init} of Figure~\ref{alg:main}.\\
    Parameter: $\textsf{timeOut}$.\\
  \textbf{PreProcess}: Each participant $P_i$ performs the following steps.
  $\pi$ is a parameter corresponding to the number of signing operations that can be performed before doing another pre-process step.
  \begin{enumerate}
      \item Create an empty list $L_i$. For $1\le j \le \pi$ do:
      \begin{enumerate}
          \item Sample single-use nonces $(d_{ij},e_{ij})\xleftarrow{\$}\mathbb{Z}_q^*\times \mathbb{Z}_q^*$.
          \item Derive commitment shares $(D_{ij},E_{ij}) = ({d_{ij}G},{e_{ij}}G)$.
          \item Append $(D_{ij},E_{ij})$ to $L_i$.
          Store $((d_{ij},D_{ij}),(e_{ij},E_{ij}))$ for later use in signing operations.
      \end{enumerate}
      \item Publish $(i,L_i)$ to the PoS blockchain.
  \end{enumerate}
  
  $\textbf{Sign}(m)$\\
  Each participant $P_i$ does the following:
  \begin{enumerate}
  \item Compute S, the set of $t$ participants for signing using $\RB_i$ as follows:
 \begin{enumerate}
     \item Compute $H(id||\RB)$.
     \item The smallest $t$ hashes are the id selected for signing.
 \end{enumerate}
      \item  Fetch the next available commitment for each participant $P_i\in S$ from $L_i$ and construct $B =  \langle (i,D_i,E_i) \rangle_{i\in S}$.
      \item Compute the set of binding values
      $\rho_l = H_1(l,m,B),l\in S$ and derives the group commitment $R = \sum_{l\in S}( D_l + {\rho_l}E_l)$
      and the challenge $c = H_2(m||R||Q)$.
      \item Each $P_i\in S$ computes their response using their secret share $s_i$ by computing $z_i = d_i +(e_i\cdot \rho_i)+\lambda_i\cdot s_i\cdot c$ using $S$ to determine the $i^{th}$ Lagrange coefficient $\lambda_i$ as follows:
      if $S= \{P_{i_1}\cdots,P_{i_t}\}$ represents the participants identifiers then $\lambda_i = 
      \prod_{j\in\{i_1,\dots,i_t\},j\ne i}
      \frac{P_j}{P_j-P_i}$.
      \item Each $P_i$ securely deletes $(d_i,e_i)$ from their local storage and then post $z_i$ to the PoS chain.
      \item After all the shares from participants in S are included in the PoS chain, each participant performs the following steps:
      \begin{enumerate}
          \item Derive  $R=\sum_{i\in S}R_i$ and $c = H_2(m||R||Q)$.
          \item Verify that ${z_i}G\stackrel{?}{=} R_i+ ({c\cdot \lambda_i})\cdot Y_i$ for each signing share $z_i,i\in S$. 
          
          If it fails, report the misbehaving participant(s) by publishing a message on the PoS blockchain with the proof of misbehaviour(s) (i.e.,
          ${z_i}G$ and $ R_i + ({c\cdot \lambda_i}) Y_i$ for each cheating player)
          and abort.
          \item If no participants was misbehaving, compute $z = \sum_{i\in S} z_i$.
          \item Compute $\sigma =(z,R)$ to the PoS blockchain.
      \end{enumerate}
\item If after $\textsf{timeOut}$ blocks since the begining of the protocol, some shares have not been posted to the PoS chain, abort the protocol and add the corresponding participants in the list of misbehaving players.
  \end{enumerate}
  Output: $(\sigma,H)$ if the protocol completed, $(\textsf{abort},S')$ else, where $S'$ is the set of players who have not misbehaved during the execution of the protocol.

}}}
 
    \caption{Signing Algorithm}
    \label{alg:signing}
\end{figure*}

\begin{algorithm*}
\begin{algorithmic}[1]
{\scriptsize
\Import
PoS \EndImport
\Import
PoS.PowerTable as PT
\EndImport
\Import
BTC
\EndImport
\Import
PrivateMessage as PM
\EndImport
\Import IPFS \EndImport
\Import{Signing Algorithm (Algorithm~\ref{pseudocode:signing}), Distributed Key Generation Algorithm (Algorithm~\ref{pseudocode:dkg}) }\EndImport
\Param
\State $id$ \Comment{The node id}
 \State $u$ \Comment{Tolerated difference between local configuration and current configuration}
 \State f \Comment{ Fault tolerance of the current configuration}
 \State $Y$ \Comment{Number of blocks to wait for the DKG to complete}
\EndParam
\Init 
\State $C_{cur}\gets C_0$ \Comment{Current configuration}
\State $C_{last}\gets C_0$ \Comment{Last configuration}
\State $pk_{cur}\gets \pk_0$
\Comment{Initial public key}
\State CurrentShares $\gets$ empty dictionary
\Comment{Share of the aggregated public key}
\State $S_0\gets C_{cur}$ \Comment{ Non-misbehaving participants}
\State $\text{misbehavingPlayers}\gets\emptyset$\Comment{Set of misbehaving participants in the DKG}
\State $i\gets C_{cur}.members[id].getIndex()$ \Comment{Node's index}
\State $\tx_{last}\gets\tx_0$ \Comment{Initial transaction as defined in Section~\ref{sec:init}}
\EndInit
  \Event{receiving $PT.\text{update}(req) \wedge C_{last}\triangle C_{cur}<u$}\Comment{Configuration request received}
\If{ $req = \langle p,``join"\rangle$}
  \State $C_{cur}.members \gets C_{cur}.members\cup \{p\} $
  \EndIf
  \If{ $req = \langle p,``leave"\rangle$}
  \State $C_{cur}.members \gets C_{cur}.members\setminus \{p\} $
  \EndIf
  \EndEvent
 \Event{$C_{last}\triangle C_{cur}\ge u$
\Comment{After $u$ (un)registrations} }
 \State $X\gets $ PoS.CurrentBlock()
 \State Do Algorithm~\ref{pseudocode:dkg} (DKG)
\EndEvent
\Event{ PoS.CurrentHeight == PoS.Height(X)+Y}\Comment{Give enough time for the DKG to complete}
  \State $S_0\gets C_{cur}.getIndexes()\setminus $misbehavingPlayers \Comment{set of indexes of non-cheating players in the DKG}
 \State $pk_{new}\gets \sum_{j\in S_0}\text{CurrentShares}[j]$ \Comment{Compute the aggregated key}
 \State $s_i\gets \sum_{j\in S_0} s_j^i$  \Comment{Compute the share of the secret key}
  \State $\ckpt\gets PoS.Blockhash(X)$
  \State $q\gets \pk_{new} + H_{TapTweak}(\pk_{new}||\ckpt)G$ \Comment{Taproot address}
  \State $o\gets 0$ \Comment{Counter for the pre-process step} 
 \If{BTC.latestCheckpoint.UTXO $\ne \tx_{last}$} \Comment{Check the Bitcoin blockchain for the previous checkpointing transaction}
 \State BTC.Broadcast($\tx_{last}$) \Comment{Send latest checkpoint}
 \EndIf
 \State IPFS.push($C_{cur}$)
\State $cid\gets
IPFS.getCid(C_{cur})$
  \If{$id \in C_{last}$} \Comment{Members associated with $\pk_{cur}$ sign $\tx$}
\State\label{line:signing} do Algorithm~\ref{pseudocode:signing}\Comment{Signing protocol with other members}

 \State $C_{last}\gets C_{cur}$
  \State $\pk_{cur}\gets q$
  \State CurrentShares $\gets\emptyset$

\EndIf
 \EndEvent

}
\end{algorithmic}
\caption{Main algorithm}\label{pseudocode:main}
\end{algorithm*}

\begin{algorithm*}
\begin{algorithmic}[1]
{\scriptsize
\Import{MainAlgorithm}\EndImport
\Param
\State $t \gets 0.5|C_{cur}|+1$ \Comment{Number of parties controlled by the adversary}

\EndParam
 \If{ $id \in C_{cur}$} \Comment{Only member of the new configuration perform the DKG}
 \State Timeout.Start() \Comment{We use a timeout to detect aborting players (this can be in terms of blocks in the PoS chain)}
 \State $r_i\xleftarrow{\$} \mathbb{Z}_q$
; $a_{i0}\gets r_i$
\For{$k\in\{1,\dots,t-1\}$}
\State $a_{ik}\xleftarrow{\$} \mathbb{Z}_q$ \EndFor 
 \State  $f_i(u) \gets \sum_{k=0}^{t-1}a_{ik}u^k$
 \For{$j\in\{1,\dots,|C_{cur}.members|\}$}
 \State PM($\langle \text{SHARE},s^j_i=f_i(j)\rangle,C_{cur}.members.index[j]$) \Comment{Send share of secret to each player}
 \EndFor
 \EndIf
  \If{id$\in C_{cur}$}
\For{$k\in\{0,\dots,t-1\}$}
$A_{ik}\gets a_{ik}G$
\EndFor
\State PoS.Broadcast($\langle \text{secretCommitments}, A_{i0},\dots,A_{i(t-1)}\rangle_i$)
\EndIf
\Event{$\langle \text{SHARE}, s_j^i\rangle_j$ received for all $j$}
\State Timeout.Restart() \Comment{All secret shares were received}
\EndEvent
\Event{TimeOut.Done()}
\For{$j\in\{1,\dots,|C_{cur}.members|\}$}
 \If{($\langle \text{SHARE}, s_j^i\rangle)_j$== nil }\Comment{Some party did not send their private share}
  \State misbehavingPlayers.append$(j)$ \Comment{Add aborting players to list of misbehaving participants}
 \EndIf
 \EndFor
 \State Timeout.Restart()
\EndEvent
 \Event{PoS.Receive($\langle \text{secretCommitments},A_{j0},\dots,A_{j(t-1)}\rangle_j$)  }
 \If{$j\in\{1,\dots,|C_{cur}.members|\}$ and $s^j_i \ne \sum_{k=0}^{t-1}i^kA_{jk}$}
 \State misbehavingPlayers.append($j$)
 \Else
 \State CurrentShares.append($j,A_{j0}$)
  \EndIf
 \EndEvent
 \Event{PoS.Receive($\langle \text{secretCommitments},A_{j0},\dots,A_{j(t-1)}\rangle_j$) for all $j$ or Timeout.Done()}\Comment{All commitments were received or timeout expired}
  \If{PoS.Read($\langle \text{secretCommitments}\rangle)_j$== nil }
  \State misbehavingPlayers.append$(j)$ \Comment{Add aborting players to list of misbehaving participants}
 \EndIf
 \State v = $\emptyset$
 \For{$j\in\{1,\dots,|C_{cur}.members|$}
 \If{$j \in \text{misbehavingPlayers}$}
 v.append($f_j(i)$)
 \Else \
 v.append(NoComplaint)
 \EndIf
 \EndFor
\State PoS.Broadcast($\langle$complaintSecret, v$ \rangle$) \Comment{Send a list of (potentially empty) complaints}
\State Timeout.Restart()
 \EndEvent
 \Event{PoS.Receive($\langle$complaintSecret,v $\rangle_k$)}
 \For{$j\in\{1,\dots,|C_{cur}.members|$} \Comment{Receive other parties' list of complaints}
\If{ v[j]$\ne$ NoComplaint}
\ misbehavingPlayers.append$(j)$\Comment{Add complaints against $j$}
\EndIf
\EndFor
\If{v[i]$\ne$ NoComplaint} PoS.Broadcast($\langle \text{complaintAnswer},s^i_j,i\rangle$) \Comment{Reply to complaint against self}
\EndIf
 \EndEvent
 \Event{PoS.Receive($\langle$complaintAnswer, proof, $j\rangle_l$)} \Comment{$j$ can answer a complaint from $l$}
 \State $s^j_l\gets $Parse(proof)
 \State $(A_{jk})_{k=0}^{t-1}\gets PoS.Read(\langle \text{SecretCommitments}\rangle_j)$
 \Comment{Get $j$ commitments}
 \If{$s^j_l G == \sum_{k=0}^{t-1} l^k A_{jk}$}
 \State misbehavingPlayers.remove$(j)$
 \EndIf
 \EndEvent
  \Event{PoS.Receive($\langle \text{complaintSecret}\rangle_j$) for all $j\wedge$ misbehavingPlayers == $\emptyset$}\Comment{All complaints (and potentially answers) were received}
 \State return \Comment{Finish the protocol}
 \EndEvent
 \Event{Timeout.Done()} \Comment{Leave enough time for answers to be received}
\State return
 \EndEvent
}
\end{algorithmic}
\caption{Distributed Key Generation}\label{pseudocode:dkg}
\end{algorithm*}

\begin{algorithm*}
\begin{algorithmic}[1]
{\scriptsize
\Import{MainAlgorithm}\EndImport
\Param $\pi$ \Comment{Number of pre-process steps} \EndParam
\State Timeout.start()
\State $L_i\gets \emptyset$
\State $v\gets$ PoS.Height(X)
\State $RB\gets RB_v$
\For{$j\in \{0,\dots,\pi\}$}
\State $(d_{ij},e_{ij})\xleftarrow{\$}\mathbb{Z}_q^*\times \mathbb{Z}_q^*$
\State $(D_{ij},E_{ij}) = ({d_{ij}}G,{e_{ij}}G)$
\State $L_i.append(D_{ij},E_{ij})$
\EndFor
\State PoS.Broadcast($\langle \text{PreProcess, }i, L_i\rangle$)
\State $B\gets \emptyset$
\State $S'\gets \emptyset$
\For{$i\in C_{last}$}
\State $S'\gets S'\cup( H(RB|| i.ID))$ \Comment{Pseudo-randomly choose set of signers}
\EndFor
\State $S'\gets order(S')$
\State $S\gets S'[:f|C_{last}|+1]$
\Comment{Choose t+1 participants for signing}
\For{$k\in S$ }\label{psuedocode-line:signingstart}
\State $(D_{ko},E_{ko})\gets$PoS.Read($\langle \text{PreProcess},k, L_k[o]\rangle)$
\State $B.append((k,D_{ko},E_{ko}))$
\EndFor

\For{$l\in S$}
\State $\tx\gets BTC.TX(\pk_{cur}\rightarrow(all,q),(0,OP_{RETURN}=cid ))$ \Comment{Compute the transaction}
\State $\rho_l\gets H_1(l,\tx,B) $ for $l\in S$
\State $\lambda_i\gets\prod_{j\in S,j\ne i}  \frac{p_j}{p_j-p_i}$
where $p_j$ is the identifier of participant $j$
\State  $R \gets \sum_{l\in S} D_{lo} + {\rho_l}E_{lo}$, $c \gets H_2(\tx ||R||q)$
\State $z_i \gets d_{io} +(e_{io}\cdot \rho_i)+\lambda_i\cdot s_i\cdot c$ 
\State delete    $((d_{io},D_{io}),(e_{io},E_{io}))$ from local storage 
\State PoS.Broadcast($\langle\text{SHARE}, z_i\rangle$)
\EndFor
\If{$id \in S$}
\Event{PoS.Receive($\langle\text{SHARE}, z_k\rangle$) from all $k\in S$}\Comment{All shares were received}
\State CheatingPlayers$\gets \emptyset$ 
\For{$k\in S$}
\State $Y_k= \sum_{j\in S_0}\sum_{w=0}^{t-1}k^wA_{jw}$
\State $\rho_k \gets H_1(k,\tx,B)$, $R_k \gets D_{ko}+ {\rho_k}E_{ko}$,  $R\gets \sum_{k\in S}R_k$,
$c \gets H_2(\tx ||R||q)$
\If{ $g^{z_k}\ne R_k+ {c\cdot \lambda_k}\cdot Y_k$ }
       \State CheatingPlayers.append($k$)
\EndIf
\EndFor
\If{CheatingPlayers$\ne \emptyset$}
\State PoS.Broadcast($\langle\text{RESTART SIGNING, CheatingPlayers}\rangle$)
\State restofplayers $ \gets C_{cur}.getIndexes() \setminus S $
\State $S\gets S\setminus$CheatingPlayers
\State S.append(restofplayers[:$|\text{CheatingPlayers}|$]) \Comment{add as many players as were removed}
\State $o\gets o+1$
\State Timeout.Restart()
\State go to line ~\ref{psuedocode-line:signingstart}
\Else
\State  $z \gets \sum_{i\in S} z_i$

       \State $c\gets PoS.Blockhash(X)$
       \Comment{Commitment to the blockchain}
      \State $\sigma'\gets \sigma + H(\tx||R||q) H(\pk_{cur}||c)$ \Comment{compute taproot signature}
       \State BTC.Broadcast(tx, $\sigma'$)
       \State PoS.Broadcast(tx,$\sigma'$)
    \State return
\EndIf
\EndEvent
\Event{Timeout.done()}\Comment{We implement a timeout to deal with aborting participants}
\For{$p\in S$}
\If{PoS.Read$(\langle\text{SHARE}\rangle_p )== nil$}\Comment{p hasn't submitted its share}
\State CheatingPlayers.append(p)
\Else
\State $\rho_k \gets H_1(k,m,B)$, $R_k \gets D_{kj}+ {\rho_k}E_{kj}$,  $R\gets \sum_{k\in S}R_k$,
$c \gets H_2(\tx||R||q)$
\If{ $g^{z_k}\ne R_k+ {c\cdot \lambda_k}\cdot Y_k$ }
\State  CheatingPlayers.append(p)
\EndIf
\EndIf

\EndFor
\State PoS.Broadcast($\langle\text{RESTART SIGNING, CheatingPlayers}\rangle$)
\State restofplayers $ \gets C_{cur}.getIndexes \setminus S $
\State $S\gets S\setminus$CheatingPlayers
\State S.append(restofplayers[:$|\text{CheatingPlayers}|$]) \Comment{add as many players as were removed}
\State $o\gets o+1$
\State go to line~\ref{psuedocode-line:signingstart}
\EndEvent
\EndIf
}
\end{algorithmic}
\caption{Signing algorithm}\label{pseudocode:signing}
\end{algorithm*}

\begin{algorithm*}
\begin{algorithmic}[1]
{\scriptsize
\Import BTC \EndImport
\Import IPFS \EndImport
\Import PoS \EndImport
\Param
$\pk_0$ \Comment{Initial public key}
\EndParam
\State $\tx_0\gets$  BTC.output($\pk_0$) \Comment{In the case of multiple transactions spent by $\pk_0$, choose the first one}
\State $i\gets 0$
\While{output is unspent}
\State output$\gets$BTC.getOutput($\tx_i$) \Comment{Get chain of transactions}
\State $i\gets i+1$
\EndWhile
\State cid $\gets$ output.$OP_{RETURN}$
\State $Q\gets$ output.TaprootAddress
\State $members\gets$ IPFS.getData(cid) \Comment{Get the configuration from IPFS}
\For{m in $members$}
\State PoS$\gets$query(m,PoS) \Comment{get the latest PoS state from the current members}
\EndFor
\State $c\gets$ PoS.getLatetsCheckpoint \Comment{verify checkpoint}
\State $\pk\gets$ PoS.getLatestAggregatedKey
\If{$Q == \pk + H_{Taproot}(\pk||c)G$ }\label{code:check-checkpoint}\Comment{Verify that the state of the database is consistent with the Bitcoin checkpoint}
\State return 1 
\Else
\State PoS $\gets$ PoS.RemoveBlocks(after c) \Comment{Roll back the PoS chain to the previous checkpoint}
\State $c\gets $PoS.getLatetsCheckpoint
\State $\pk\gets$ PoS.getLatestAggregatedKey
\State Go to step~\ref{code:check-checkpoint}
\EndIf
}
\end{algorithmic}
\caption{Verification}\label{pseudocode:verif}
\end{algorithm*}

\section{Security argument}\label{sec:security}
In this section we present the arguments for why our protocol is secure.
We need to prove two things: (1) that any checkpoint pushed onto the Bitcoin blockchain is \emph{correct}, i.e., that it corresponds to the valid state of the PoS (according to honest online validators);
(2) that checkpoints will be pushed regularly.
These two properties correspond, loosely, to the safety and liveness properties of our scheme.

\subsection{Safety}


We consider the following statement, which we prove by induction, for $k\in\mathbb{N}$:
\textit{An adversary as defined in Section~\ref{sec:assumptions} cannot create any incorrect checkpointing transaction $\tx^A_i$ for any $0\le i\le k$ such that $\tx^A_i$ will be accepted by an honest verifier that follows the verification algorithm as defined in Section~\ref{sec:verify}}.
An incorrect checkpoint transaction is a transaction that contains a commitment to an incorrect chain (i.e., a chain created as part of a \LRA).

\paragraph{Base Case}
First, we show that the adversary cannot create an alternative initial transaction $\tx_0$.
At the time where the initial transaction is created, the adversary controls at most $t_0$ participants (assumption 2) and hence, by security of the DKG and signing algorithms, cannot unilaterally sign a transaction coming from $Q_0$.
After $L$ configurations, the adversary do obtain all the keys from $C_0$ and is able to create transaction coming out from this address, however, this happens after height $h_0$ on the Bitcoin blockchain by assumption 3 and hence any transaction sent by the adversary from $Q_0$ will not be accepted by any verifier according to our verification algorithm presented in Section~\ref{sec:verify} step~\ref{enum:verif-init}.
Hence the adversary cannot create an initial checkpoint transaction that will be accepted by any verifier.

\paragraph{Induction step}
Let's assume that our statement is true for $k-1$, i.e., the adversary cannot create any incorrect checkpointing transaction up to $k-1$ (i.e., $\tx^A_0,\dots,\tx^A_{k-1}$).
We show that our statement is then also true for $k$.
It is enough to show that the adversary cannot create any incorrect checkpointing transaction $\tx^A_{k}$.
Let's denote $i$ the current configuration number (i.e., according to online validators).
There are two cases to consider.
The first case is the case where $k<i-L$. Then by assumptions the adversary has all the keys associated with $Q_{k}$ (assumption 1) and a transaction $\tx_k$ that spent $\tx_{k-1}$ has already been included in the blockchain (assumption 3).
Because $\tx_{k-1}$ has already been spent, $\tx^A_k$ cannot include $\tx_{k-1}$ in its inputs (as an input can only be spent once according to Bitcoin's rules).
Moreover by induction assumption there is no other transaction $\tx^A_{k-1}$ to be included as an input to $\tx^A_k$ that the adversary could create that would be accepted by any verifier.
Hence, according to our verification algorithm step~\ref{enum:verif-induction} $\tx^A_k$ will not be accepted by any verifier.

The second case is the case where $k\ge i-L$. In this scenario, it could be the case that transaction $\tx_{k-1}$ is still unspent. 
By design the only spendable outputs of $\tx_{k-1}$ is $Q_k$.
However, according to assumption 2, the adversary only holds a fraction $f$ of configuration $k$ and hence cannot create a transaction that is spent by $Q_k$ and cannot spent transaction $\tx_{k-1}$.

\subsection{Liveness}
The reasons why an adversary cannot stop the signing from going ahead and the checkpoints from happening are as follows.
(1) The robustness of the DKG ensures that an adversary cannot stop the rest of the players from computing an aggregated public key.
(2) The adversary could delay the signing process by aborting; however, aborting or misbehaving players will be detected and excluded from the signing in the next iteration.
(3) The assumption about the stability across configurations ensures that enough honest participants will be able to perform the signing, i.e., we assume that enough participants from each configuration will remain available in the system long enough to sign and give the signer power to the next configuration.

\begin{figure*}[h]
\centering
     \begin{subfigure}[b]{0.48\textwidth}
         \centering
  \includegraphics[width = \linewidth]{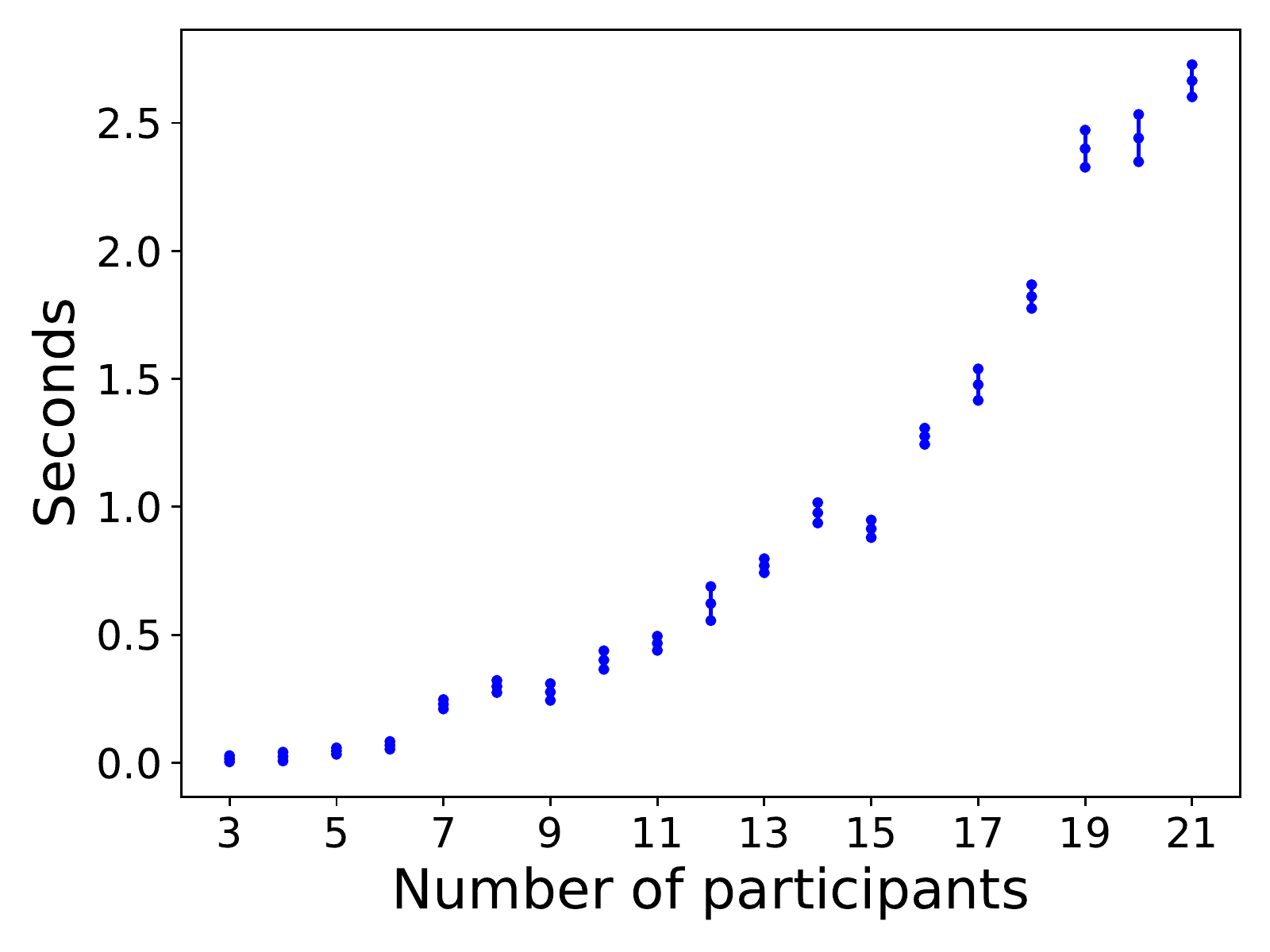}
\caption{DKG with every participants honest.}
\label{fig:dkg}
     \end{subfigure}
     \hfill
     \begin{subfigure}[b]{0.48\textwidth}
         \centering
  \includegraphics[width = \linewidth]{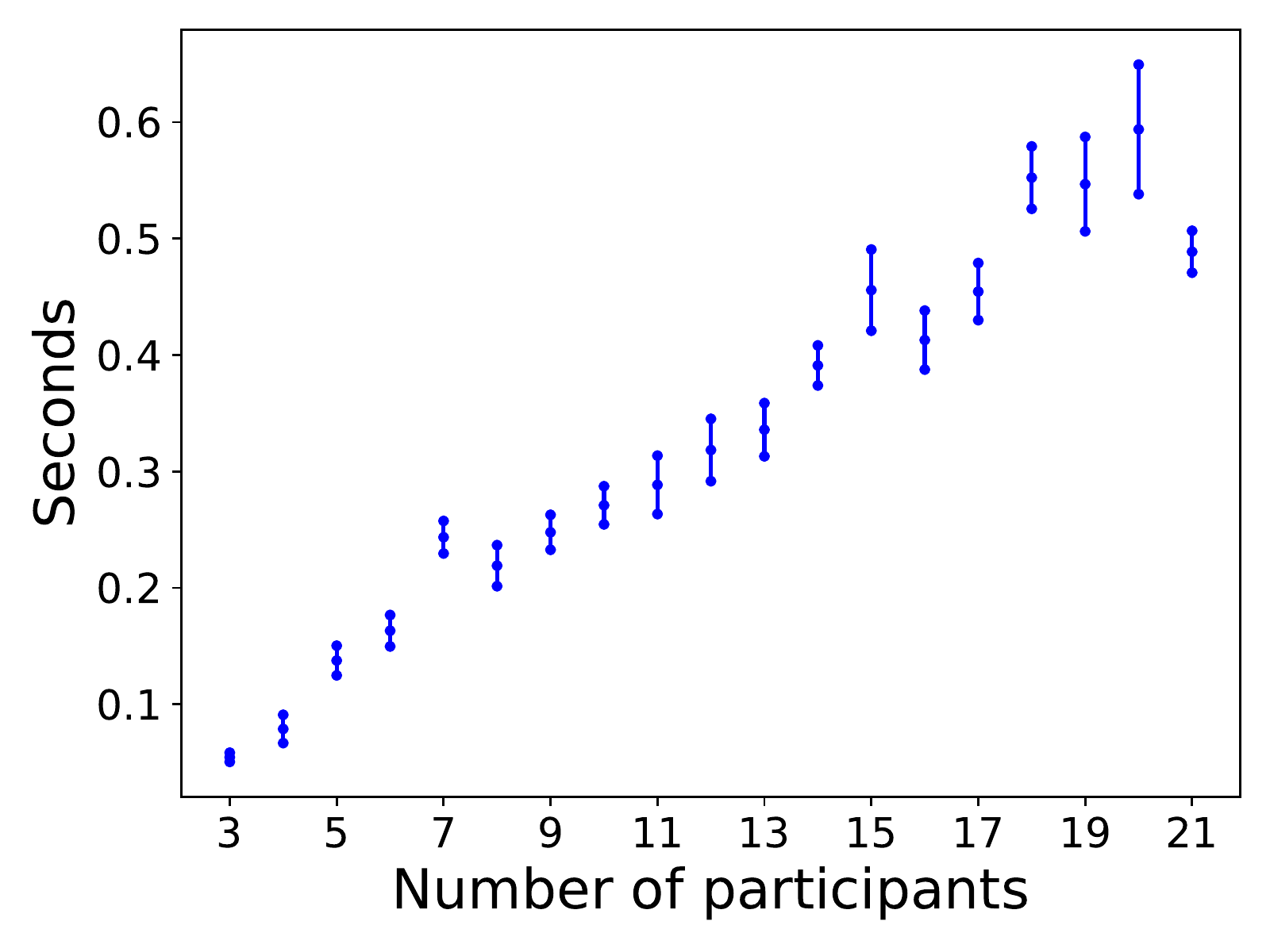}
\caption{Signing with every participants honest.}
\label{fig:signing}
     \end{subfigure}
\caption{Execution time of our DKG and Signing Implementation. Each vertical bar represent the confidence interval of the execution time as seen by each different node.}
\label{fig:experiments}
\end{figure*}

\section{Implementation and Evaluation}\label{sec:implementation}

We implement the protocol from Section~\ref{sec:protocol}
using the Go Programming Language.
For the underlying PoS chain, we forked the open-source Eudico framework,\footnote{\url{https://github.com/filecoin-project/eudico}} developed by Protocol Labs, that provides a delegated Proof-of-stake consensus protocol option.
We used a simplified version of this, where only one PoS miners creates blocks, as this does not impact our experiments.
We used an open-source library developed by the Taurus group\footnote{\url{https://github.com/taurusgroup/multi-party-sig}} for the DKG and signing, that we adapted for our needs and used both Bitcoin regtest and testnet for our experiments.
For storing the data associated with each configuration, we implemented a key-value database, maintained by the PoS validators on top of the PoS chain.

The code is open source.\footnote{\url{https://github.com/filecoin-project/eudico/tree/B2-bitcoin-checkpointing}}
We run the experiments on a single virtual machine (32 GB RAM, 8 vCPUs, 640 GB SSD) on Amazon Lightsail using a Kubernetes deployment.

We implemented the verification process, however we did not include any metrics in this paper as this was tested only on the Bitcoin Testnet and may not be representative of the mainnet. 

We measure
the execution times of the DKG and the signing protocol in Figure~\ref{fig:experiments}. We only included the case where everyone cooperates in our graphs as in the case of failures our protocol relies on a timeout (to detect aborts) hence the execution time of the protocol with failures is constant and only depends on the timeout chosen.
\ifproceeding\else{We included the graphs of the execution time in the event of failures in the Appendix for reference.}\fi
While the number of validators in a PoS protocol varies depending on a particular blockchain system, we show results with up to 21 validators, which corresponds to the number of validators in a delegated PoS such as EOSIO~\cite{eosio}, where 21 validators are elected on a rotating basis to run the consensus protocol.
In Figure~\ref{fig:experiments}, we plot the confidence interval of the execution time of the DKG and signing protocol sampled over all the participating nodes and repeated a dozen times.

We notice in our graph that the signing scales better than the DKG as it increases from less than 0.1 second with 3 participants to around 0.6 second with 21 participants whereas the DKG goes up to above 2.5 seconds with 21 participants.
This is expected as the signing only requires 2 broadcast messages per participants (the pre-process and the share of the signature) whereas the DKG requires private messages between every participants as well as broadcast messages.

\section{Related Work}\label{sec:related}
\LRA\ have long been studied in the field of PoS and
other types of checkpointing have been proposed that either rely on some sort of central authority~\cite{ppcoin} or on additional assumptions~\cite{azouvi2020winkle}. 
Like the solution from Steinhoff et al.~\cite{steinhoff2021bms}, this paper offers a fully decentralized solution without additional security assumptions (as in~\cite{azouvi2020winkle}) other than the ones needed for the security of the underlying PoS. 

Kuznetsov and Tolkih propose an alternative solution to addressing long-range attacks in BFT/PoS \cite{KuznetsovT20}, using forward-secure digital signatures. However, this solution is inapplicable in the rational adversary model, in which rational nodes might simply not follow the assumptions of forward-secure digital signatures, retaining their old private keys to mount attacks in the future.

Babylon~\cite{tas2022babylon} was proposed concurrently to our work and is a defense against \LRA\ that is also based on leveraging the security guarantees provided from Bitcoin's Proof-of-work.
In this work, every PoS miner can post a checkpointing transaction into the Bitcoin blockchain which then acts as a timestamping mechanism and thus thwarts \LRA. 
Whenever a block is mined on the PoS chain, PoS validators can submit a commitment for this block, and this commitment is included in the Bitcoin PoW chain.
In the case of two conflicting blocks in the PoS chain,  the one whose commitment was submitted first in the Bitcoin chain is chosen by the fork-choice rule.
Babylon goes further and also protects the underlying PoS chain against super-majority and censorship attacks.
Their scheme is more scalable than Pikachu, as it does not require any additional threshold signing. On the other hand, since the checkpointing transactions on the Bitcoin blockchain are not linked together, as is the case in Pikachu, the verification algorithm is much less efficient as one would need to search exhaustively through all the Bitcoin transactions to find all the possible PoS checkpoints and ensure that they have the correct PoS chain.

Lastly, on the topic of Stake-based Threshold Multisignatures, Mithril~\cite{chaidos2021mithril} and Dfinity~\cite{groth2021non} both propose scalable and efficient schemes that are however not compatible with Bitcoin and could thus not be used in the context of checkpointing onto Bitcoin.

\section{Conclusion}\label{sec:open-problems}
We presented a checkpointing mechanism designed to secure PoS blockchains by leveraging the security guarantees provided by Bitcoin's PoW.
Our protocol uses Taproot, allowing for the checkpoints to be constant in the size of PoS validators and indistinguishable from any other Taproot's transaction.
We implemented a PoC for our protocol and measured its efficiency. 
The main issue of our approach is that it does not scale well.
This is especially true if we consider a flat model where each unit of power corresponds to a different public key; we could easily end up dealing with tens of thousands of keys, even when the number of actual participants is much smaller, greatly increasing the latency of the protocol.
Although some techniques such as sampling~\cite{chaidos2021mithril} or ad-hoc threshold multi-signature schemes~\cite{gavzi2019proof} have been proposed to help scale weighted threshold signature schemes, those techniques are not currently compatible with Bitcoin's spending rules.

Another problem left for future work is that of fully incentivising the participation in the protocol, which we started doing in Section~\ref{sec:init}.

\section*{Acknowledgments}
The authors would like to thank Nicolas Gailly and Rosario Gennaro for very useful discussions about this paper.

\bibliographystyle{abbrv}
\bibliography{ref}

\ifproceeding\else{
\appendix

\section{Extra Figure}
\begin{figure*}[h]
\centering
     \begin{subfigure}[b]{0.48\textwidth}
         \centering
  \includegraphics[width = \linewidth]{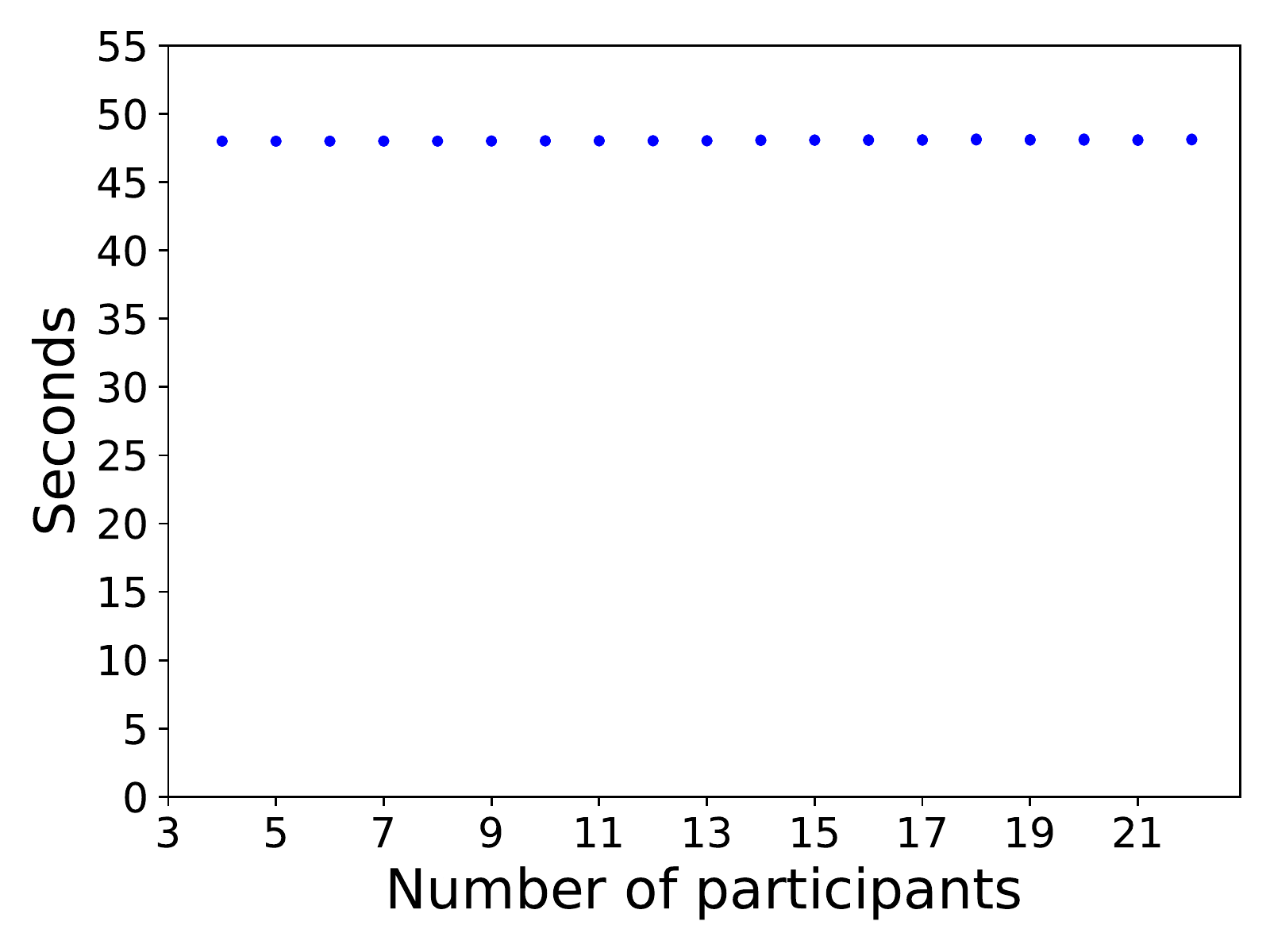}
\caption{DKG.}
\label{fig:dkg}
     \end{subfigure}
     \hfill
     \begin{subfigure}[b]{0.48\textwidth}
         \centering
  \includegraphics[width = \linewidth]{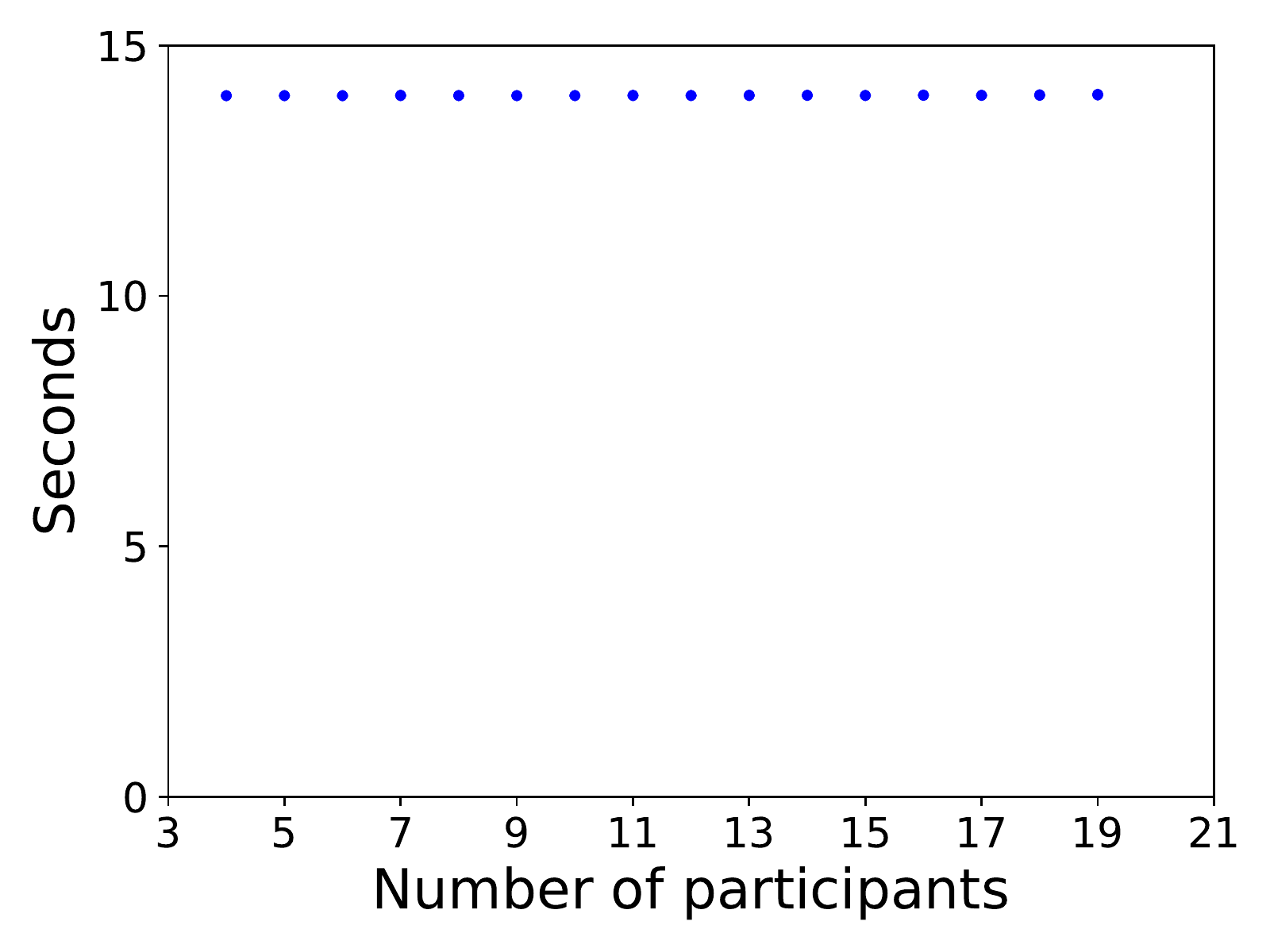}
\caption{Signing.}
\label{fig:signing}
     \end{subfigure}
\caption{Execution time of our protocols with failures (here an aborting participant). A (conservative) timeout of 12 seconds was chosen. In the DKG case, the timeout is repeated over the different rounds (waiting for shares of the secrets, for public commitments, for complaints, for complaint answers). This explains the much longer execution time.}
\label{fig:experiments}
\end{figure*}

}\fi

\end{document}